%% file: main.tex
\documentclass[conference]{IEEEtran}

\usepackage{booktabs} 
\usepackage{enumitem}
\usepackage{graphicx}
\usepackage{makecell}
\usepackage{listings}
\usepackage{color}
\usepackage{verbatim}

\usepackage{algorithmicx}
\usepackage{algorithm}
\usepackage[noend]{algpseudocode}

\algnewcommand\INPUT{\item[{\textbf{Input:}}]}
\algnewcommand\OUTPUT{\item[{\textbf{Output:}}]}

\definecolor{dkgreen}{rgb}{0,0.6,0}
\definecolor{gray}{rgb}{0.5,0.5,0.5}
\definecolor{mauve}{rgb}{0.58,0,0.82}

\newcommand{\newspace}{\vspace*{3pt}}

\newcommand{\aaf}{\vspace*{-6pt}}
\newcommand{\af}{\vspace*{-3pt}}

\graphicspath{ {images/} }

\usepackage{cite}
\usepackage{amsmath,amssymb,amsfonts}
\usepackage{graphicx}
\usepackage{textcomp}
\usepackage{xcolor}

\usepackage[hidelinks]{hyperref}

\begin{document}

\title{Code-less Patching for Heap Vulnerabilities Using Targeted Calling Context Encoding}

\author{\IEEEauthorblockN{Qiang Zeng$^{\dag}$, Golam Kayas$^{\ddag}$,  Emil Mohammed$^{\ddag}$, Lannan Luo$^{\dag}$, Xiaojiang Du$^{\ddag}$, and Junghwan Rhee$^{\S}$}
\IEEEauthorblockA{$^{\dag}$\textit{University of South Carolina } 
\qquad
$^{\ddag}$\textit{Temple University} \qquad
$^{\S}$\textit{NEC Lab} \\
\{zeng1,lluo\}@cse.sc.edu,  \{golamkayas,tuf58189,dux\}@temple.edu, \{rhee\}@nec-labs.com }
}

\maketitle
\thispagestyle{plain}
\pagestyle{plain}


\input{abstract}


\input{introduction}

\input{related_work}
\input{problem_scope}

\input{design_implementation}
\input{implementation}

\input{evaluation}

\input{limitations}
\input{conclusions}




\section*{Acknowledgment}
This project was supported by NSF CNS-1815144 and NSF CNS-1856380.

\bibliographystyle{IEEEtran}
\bibliography{main}

\end{document}

%% file: abstract.tex
\begin{abstract}
Exploitation of heap vulnerabilities has been on the rise, 
leading to many devastating attacks. 
Conventional heap patch generation is a lengthy procedure, 
requiring intensive manual efforts. Worse, fresh patches tend to 
harm system dependability, hence deterring users from deploying them. We 
propose a heap patching system that simultaneously has the following  prominent advantages: 
(1) \emph{generating patches without manual efforts}; 
(2) \emph{installing patches without altering the code} (so called \emph{code-less patching}); 
(3) \emph{handling various heap vulnerability types}; 
(4) \emph{imposing a very low overhead}; 
and (5) \emph{no dependency on specific heap allocators}.
As a separate contribution, 
we propose \emph{targeted calling context encoding}, which is a suite of 
algorithms for optimizing calling context encoding, an important technique with applications in many areas.
The system properly combines heavyweight offline attack analysis
with lightweight online defense generation, and provides
a new countermeasure against heap attacks. 
The evaluation shows that the system 
is effective and efficient.
\end{abstract}

%% file: introduction.tex
\section{Introduction} \label{sec:intro}
As many effective measures for protecting call stacks get deployed
(such as canaries~\cite{Cowan:1998:SAA:1267549.1267554}, reordering local variables~\cite{pax}, 
and Safe SEH~\cite{safeseh}), heap vulnerabilities gain growing attention
of attackers. Heap vulnerabilities can be exploited by attackers
to launch vicious attacks. The recent Heartbleed~\cite{heartbleed} and
WannaCry~\cite{wannacry} attacks demonstrate the dangers.
For instance, the WannaCry ransomware uses the EternalBlue exploit, which
makes use of a heap buffer overwrite
vulnerability to hijack the control flow of the victim program~\cite{wannacry}.

There are a variety of heap vulnerability types. The following
types are among the most commonly exploited types.\footnote{\emph{Double free}
\emph{was} frequently exploited; but many popular allocators,
such as the default allocator in glibc~\cite{mallopt},
have built-in double free detection now.}
(1) \textbf{Buffer overflow:} it includes both \emph{overwrite} and \emph{overread}.
By overwriting a buffer, the attack
can manipulate data adjacent to that buffer and launch
various control-data or non-control-data attacks, while
exploitation of overread can steal sensitive information 
in memory, such as address space layout and private keys.
(2) \textbf{Use after free:} it refers to accessing memory after
it has been freed. If the memory space being reused is
under the control of attackers, use-after-free bugs can 
be exploited to launch various attacks, such as control flow hijacking.
(3) \textbf{Uninitialized read:} exploitation of such vulnerabilities can
leak sensitive information.

Many approaches have been proposed to
tackle  heap vulnerabilities. 
Some systems try to discover zero-day heap vulnerabilities before software release~\cite{dowser,borg,203682}.
Yet, it is very unlikely to find all of them.
A large body of research focuses on detecting, preventing or mitigating
heap attacks (and other memory-based attacks)~\cite{addresssanitizer,
Liu2016a, Zeng:2011:CCH:1993498.1993541, DBLP:conf/ndss/TianZW0H12,
Berger:2006:DPM:1133981.1134000, Novark:2010:DSH:1866307.1866371,
Silvestro:2017:FFS:3133956.3133957, Akritidis:2010:CMA:1929820.1929836,
Younan_2015, vanderKouwe:2017:DSU:3064176.3064211, Lee_preventinguse-after-free,
Zeng2015, stepanov2015memorysanitizer, Lu:2016:UPK:2976749.2978366, milburn2017safeinit,
Nethercote:2007:VFH:1250734.1250746, Sidiroglou2005}. They usually incur a large
overhead or/and can only handle a specific type of heap vulnerabilities.
For example, MemorySanitizer~\cite{stepanov2015memorysanitizer} is 
a dynamic tool that detects \emph{uninitialized
read}; however, it incurs 2.5x of slowdown and 2x of memory overhead.
AddressSanitizer~\cite{addresssanitizer}, which detects \emph{overflows} and \emph{use after free} online, 
is deemed \emph{fast}, but still incurs 73\% slowdown and 3.4x memory overhead.
As another example, HeapTherapy~\cite{Zeng2015} proposes an
efficient \emph{heap buffer overflow} detection and response system;
however, it does \emph{not} provide methods
for detecting and handling \emph{uninitialized read} and \emph{use after free}.

When examining the
spectrum of heap security measures, we notice that handling 
heap vulnerabilities through \textbf{patching} has 
been much less studied. Patching, however, has been 
an indispensable step for handling vulnerabilities in practice.
Over decades, conventional patch generation and deployment have suffered serious
limitations. First, the patch generation is a lengthy procedure.
Even for security sensitive bugs, it takes those \emph{big}
vendors 153 days on average from vulnerability report to
patch availability~\cite{153days}. 
A study finds that only 65\% of vulnerabilities in software 
running on a typical Windows host have patches available at 
vulnerability disclosure~\cite{frei2011end}. This provides opportunities 
for attackers to exploit the unpatched vulnerabilities on a large scale~\cite{bilge2012before}.
For resource-constrained small software companies, it takes even
longer time. 
Plus, it is unlikely to
generate patches  for legacy software, whose support from the
vendor has ended. 

Second, given a vulnerability, its fresh patches 
may have not been thoroughly tested, and thus
tend to introduce stability issues and even
logic errors. Although waiting for mature patches can reduce the risk, 
it makes the exploitation window longer.
This has been a dilemma in patch deployment~\cite{timing}.

We propose a 
heap patching system that does not have the limitations
above. Our  insight is that, by changing the configuration
of heap memory allocation, \emph{all} the aforementioned
heap vulnerabilities can be addressed without altering the program code 
and, hence, no new bugs are introduced. 
Based on the configuration information,
the allocator can accordingly enhance its handling 
(i.e., allocation, initialization and deallocation) 
of buffers that are vulnerable to attacks, called \emph{vulnerable buffers},
and apply security enhancement only to them (rather than all heap buffers)
to minimize the overhead.
We thus propose to generate \emph{Heap Patches
as Configuration} and call
our system \textsc{HPaC}. 

\textsc{HPaC} consists of a \emph{heavyweight} offline patch generation
phase and a \emph{lightweight} online defense generation phase. 
In the offline patch generation phase, we use \emph{shadow memory}
to scrutinize attacks and achieve byte precision level. 
We group buffers according to their allocation-time calling contexts.
Buffers that share the same allocation-time calling context as the
buffer exploited by the attack are regarded as vulnerable buffers. 
The allocation-time calling context of vulnerable buffers along with other information
is collected to generate \emph{patches}, i.e., the configuration information. 
Next, in the online defense generation phase, the configuration 
information is loaded and the stored
calling context information
guides the allocator to recognize vulnerable buffers.
It properly combines detailed offline analysis
and highly efficient online defenses.

However, if call stack walking (as used by \texttt{gdb})
is used for obtaining calling contexts, it can incur significant slowdown, especially 
for allocation-intensive programs~\cite{pcc,pcce,deltapath}.
We thus use \emph{calling context encoding}, which \emph{continuously}
represents the current calling context in one or a few integers~\cite{pcc}.
By reading the integer(s), the encoded calling context, called 
\emph{Calling Context ID (CCID)}, can be obtained. 
By comparing the CCID for the current buffer allocation with the CCIDs stored
in the configuration information,
the online system can swiftly determine whether the new buffer is vulnerable.
Moreover, we propose \emph{targeted calling context encoding}, which
is a suite of algorithms
that can optimize many famous calling encoding
methods, such as PCC~\cite{pcc}, 
PCCE~\cite{pcce}, and DeltaPath~\cite{deltapath}.
Since calling context encoding is an important
technique with many applications, the optimization algorithms
constitute a separate contribution.

Installing
a heap patch does not change the program code. Specifically,
a heap patch is
in the form of a $\langle key, value\rangle$ tuple,
where the \emph{key} is the allocation-time CCID
of the vulnerable buffer and the \emph{value} indicates the vulnerability type and
the parameter(s) for applying the online defense.
The patches are read
into a hash table upon program initialization. It thus
takes only O(1) time to determine whether a new buffer
is vulnerable.
The online defense is enforced by intercepting
heap buffer allocation and deallocation.
Both the  hash table
initialization and the buffer allocation/deallocation
interception are \emph{transparent} to the underlying heap allocator, 
and implemented in a shared library.\footnote{In Linux,
we can load it using \texttt{LD\_PRELOAD}.} 
We thus do \emph{not}
need to change the underlying heap allocator or depend on a specific allocator.

None of the techniques used in \textsc{HPaC}, except for \emph{targeted calling context encoding}, is new. However, static analysis, code instrumentation, 
offline attack analysis, and online defense generation are
creatively combined to build a new countermeasure
against heap attacks. 
A comprehensive evaluation is performed, showing that
\textsc{HPaC} is effective and efficient.
We make the following
contributions.

\begin{itemize}

\item We properly combine heavyweight offline attack analysis and lightweight
online defense generation to build a new heap defense system
that simultaneously demonstrates the
following good properties: (1) \emph{patch generation without
manual efforts}, 
(2) \emph{code-less patching},
(3) \emph{versatile} handling of heap buffer overwrite, overread, use after free, and
uninitialized read,
(4) \emph{imposing a very small overhead}, and
(5) \emph{no dependency on specific allocators}.

\item We propose \emph{targeted calling context encoding}, a suite 
of algorithms that can optimize calling context encoding,  and 
demonstrate its application to our system.

\end{itemize}

%% file: related_work.tex
\section{Related Works}
Given the large body of research on heap memory safety, 
we do not intend to make an exhaustive list of work on the problem.
Instead, we compare \textsc{HPaC} with other automatic patch
generation techniques, and then examine critical techniques
used in our system.

\subsection{Automatic Patch/Defense Generation}
With attack inputs in hand, generating patches/defenses automatically has been a
highly desired goal.
We divide previous researches towards this goal into the following categories. 

\newspace
\noindent \textbf{Bytes pattern based signature generation.}
Given \emph{a large number} of attack inputs,
many systems (such as Honeycomb~\cite{Kreibich:2004:HCI:972374.972384},
Autograph~\cite{Kim:2004:ATA:1251375.1251394},  
and Polygraph~\cite{Newsome2005}) generate
signatures by extracting common bytes patterns from the inputs.
However, such methods usually need \emph{many} attack samples
in order to correctly mine patterns, and cannot work 
when only one or very few attack inputs are available.
False positives may be raised when benign inputs happen
to match the signatures. 
Plus, attackers can mutate the inputs to
bypass the detection.
In addition, these systems usually have deployment difficulty in
handling compressed or encrypted inputs.

\newspace
\noindent \textbf{Semantics based signature generation.}
Tools like COVERS~\cite{Liang:2005:FAG:1102120.1102150}, Hamsa~\cite{Li2006}, 
TaintCheck~\cite{Newsome05dynamictaint} and the work by Xu et al.~\cite{automatic} 
propose methods to generate semantics-based signatures; e.g., spotting the 
target system call ID used upon control flow hijacking and 
filtering out inputs that contain
that ID. They are very effective
in handling certain control flow hijacking attacks, but it is unknown
how they can be applied to addressing overread
and uninitialized read. They also have deployment difficulty in
handling compressed and encrypted attack inputs and may incur false positives. 

\newspace
\noindent \textbf{Tracking faulty instructions.}
By replaying the attacks, some systems try to pinpoint
faulty instructions that are exploited by the attacks and
try to generate patches to fix them; such systems include 
VSEF\cite{specific-taint}, Vigilant \cite{Costa:2005:VEC:1095810.1095824}, PASAN \cite{Smirnov2007} and AutoPag \cite{Lin:2007:ATA:1229285.1267001}.
A frequently employed insight is that a tainted input, e.g., due
to overwrite, should not be used to calculate the
indirect jump address. It is unknown how such systems
can handle attacks beyond control flow hijacking, e.g., 
buffer overread attacks. Plus, the
deployment of the patches requires code update, just like
conventional code patching. 

\newspace
\noindent \textbf{Trial and error  for patch generation.}
Some systems propose
genetic programming based program generation~\cite{Weimer2009b},
template based patch generation~\cite{Kim:2013:APG:2486788.2486893},
and patch generation via machine learning~\cite{long2016automatic}
to generate many patches, and test each of them against prepared
test cases until one patch
passes all the tests. However, it usually takes a lot of effort to 
prepare well-structured test cases with a decent test coverage. 
Other systems keep generating candidate patches based on certain criteria
until one can recover the program execution~\cite{rx, perkins2009automatically}. 
There is no guarantee
a qualified patch can be generated using these methods.
It is also unknown whether the qualified patch may
introduce logic errors.

While there are many works on 
automatic defense/patch generation, 
most of the proposed systems suffer one or more
of the following limitations: deployment
difficulties, false positives, requiring
many attack inputs or test cases. 
Unlike existing automatic patch generation systems,
\textsc{HPaC} supports eacy deployment without
code updates, guarantees zero false positives, 
requires only \emph{one} attack input, and handles 
multiple types of heap vulnerabilities.

\subsection{Calling Context Encoding}
\noindent\textbf{Background.} 
A \emph{calling context} is the sequence of active function calls on the call stack. 
It carries critical information about dynamic program behavior. It thus
has been widely used in debugging, testing, anomaly detection,
event logging, performance optimization, and profiling~\cite{deltapath}. For example,
logging sensitive system calls is a practice in many systems. Recording
the calling context of the system call provides important information
about the sequence of program components that gets involved and leads to the call. 

Obtaining calling contexts through stack walking is straightforward
but very expensive~\cite{pcc}. A few encoding techniques, 
which represent a calling context using one or very few
integers, have been proposed to continuously track
calling contexts with a low overhead. 
The \emph{probabilistic calling context} (PCC) technique~\cite{pcc}
computes a probabilistically unique integer ID, essentially a hash value, 
for each calling context, but does not support decoding.
\emph{Precise calling context encoding} (PCCE)~\cite{pcce}
stems from \emph{path profiling}~\cite{ball1996efficient} and supports decoding.
Another example is \emph{DeltaPath}~\cite{deltapath}, which improves PCCE 
by supporting virtual function calls and large-sized programs. 
A relevant but different problem is path encoding~\cite{ball1996efficient},
which represents program execution paths (within
a control flow graph) into integers.

Similar to \emph{targeted calling context encoding},
another work~\cite{mytkowicz2009inferred} also aims to minimize the overhead due to the encoding,
but uses a very different idea.
It performs offline-profiling runs to establish the mapping s
between stack offsets and calling contexts. It fails if the 
calling context of interest does not appear in the profiling runs.  Its reported
decoding failure rate is as high as 27\%. Finally, it does work  if variable-size 
local arrays (allowed in C/C++) are used.

\subsection{Calling Context-Sensitive Defenses}
Calling context was applied to areas \emph{beyond} debugging decades ago.
As an example, a region-based heap allocator tags
heap objects with allocation-time calling context~\cite{seidl1998segregating}.
Recently, calling context is used to generate context sensitive 
defenses~\cite{automatic,specific-taint,novark2008exterminator,Zeng2015,feng2003anomaly}.
In particular, Exterminator~\cite{novark2008exterminator} also proposes to
generate context-sensitive heap patches. 
However, our system \textsc{HPaC} differs from Exterminator 
in multiple aspects. 
(1) Exterminator performs online probabilistic attack detection (e.g., when an overflow occurs, 
it may or may \emph{not} detect it), while \textsc{HPaC} performs offline 
deterministic attack analysis and patch generation. 
How to apply patches generated by heavyweight offline analysis to lightweight
online defense generation is not trivial and solved by our work. 
(2) Exterminator does not handle overread or uninitialized-read, while \textsc{HPaC} 
handles all the frequently exploited heap vulnerability typs
including overwrite, overread, use after free, and uninitialized read. 
(3) Exterminator relies on a custom heap allocator that incurs large overheads, 
while \textsc{HPaC} does not; the defense of \textsc{HPaC} is transparent to the underlying allocator.
(4) Exterminator uses the expensive stack walking to retrieve calling contexts, 
while \emph{targeted calling context encoding} is proposed and applied in \textsc{HPaC}.
But the two works share the insight in calling context-sensitive heap patches, which we do \emph{not}
claim as our contribution.

\subsection{Shadow Memory}
Our offline heavyweight analysis makes use of shadow memory~\cite{memcheck-shadow-memory},
which tags every byte of memory used by a program 
with some information. For example, by tagging a memory
region as inaccessible, a \emph{read zone} is created.
Despite its powerful  capabilities in dynamic analysis,
it incurs very high overheads. The implementation in Memcheck, 
which is built in Valgrind,  incurs 22.2x slowdown~\cite{memcheck-shadow-memory}. AddressSanitizer significantly
improved it, but still incurs 73\% with many functionalities cut~\cite{addresssanitizer}.
Our system extends shadow memory by associating every 
heap buffer with its calling context ID. 

\vspace{5pt}
\textsc{HPaC} does not propose new techniques (except for targeted calling context encoding),
but properly combines heavyweight offline analysis (based on shadow memory)
and lightweight online defenses (based on allocation/deallocation interception and 
calling context encoding). It overcomes the challenge of applying
offline analysis results to online defenses, and carries many good
properties, such as no dependency on any custom allocator, and
handling of various heap vulnerabilities. 

%% file: problem_scope.tex
\section{Problem Statement and Architecture}
\subsection{Problem Statement}
Similar to conventional patch generation, 
our system uses collected attack inputs for 
attack investigation and patch generation.
Given a program $\mathbb{P}$ that contains a heap
vulnerability $\mathcal{V}$ and an attack input $\mathcal{I}$
that exploits $\mathcal{V}$, our system
outputs a patch $\mathcal{P}$, which, once installed,
can defeat attacks that exploit $\mathcal{V}$.
We consider the \emph{three} frequently exploited heap vulnerability types
described in Section~\ref{sec:intro}. 

But our system differs from conventional patch generation
in the following aspects. 
(1) Instead of relying on manual investigation,
patches can be generated instantly and automatically. 
(2) Rather than updating the program $\mathbb{P}$ to fix heap vulnerabilities,
patches only need to be written into a configuration file $\mathcal{C}$
to take effect.

\subsection{System Architecture}

\begin{figure}[t]
    \centering
\aaf
   \includegraphics[scale=.6]{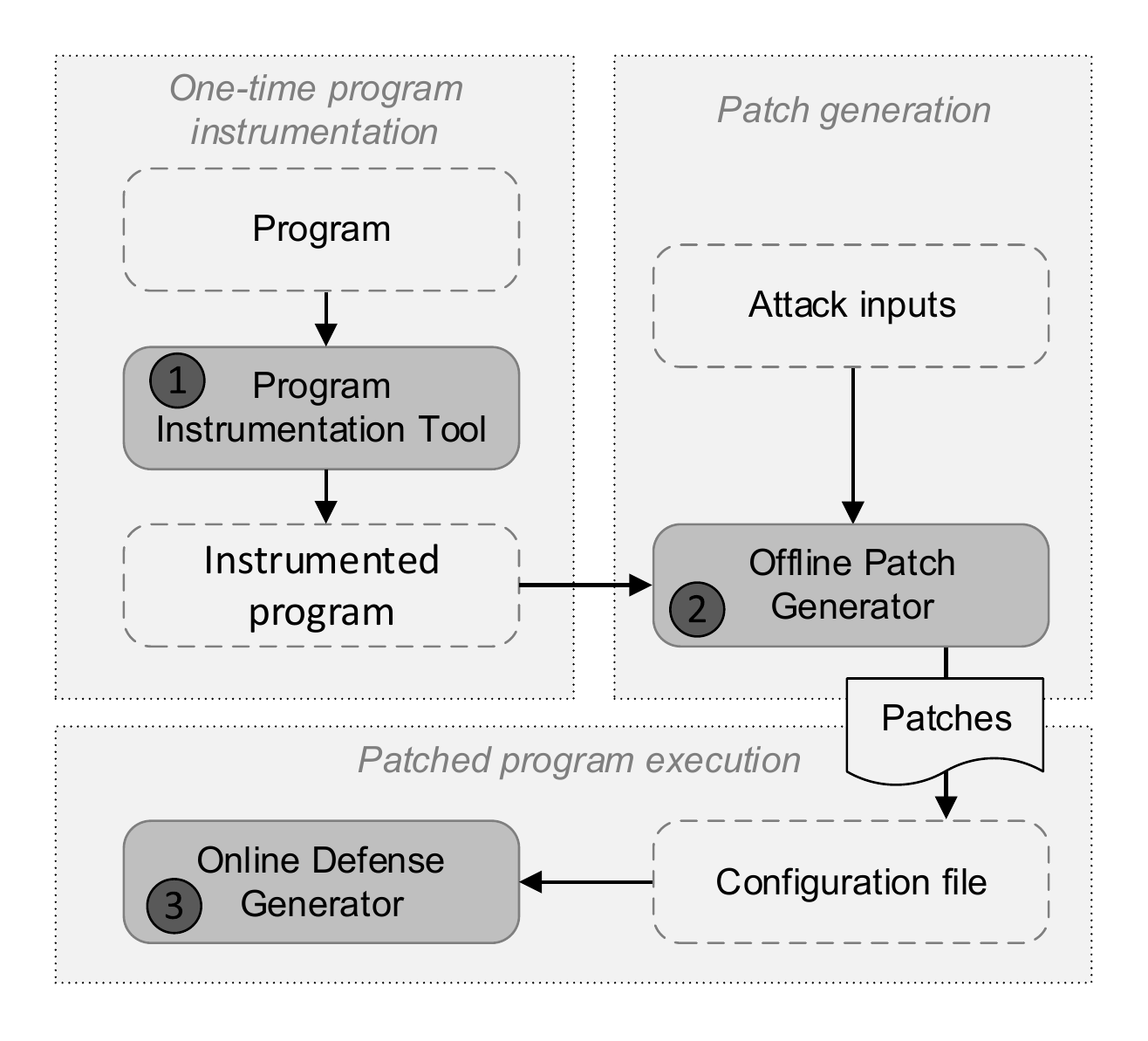}
    \aaf
    \caption{System architecture.}
    \label{fig:overview}
\af
\end{figure}

As shown in Figure~\ref{fig:overview}, the system consists of the following
components: (1) A \emph{Program Instrumentation Tool}: it builds the 
calling context encoding capability into the program (Section~\ref{sec:targeted}). 
Note that the program instrumentation is an \emph{one-time} effort. Because of the simplicity of
the instrumentation, its correctness can be verified automatically. 
The instrumented program is then
used for both offline patch generation and the online system.  
(2) An \emph{Offline Patch Generator}: it
automatically generates the patch by replaying the attack (Section~\ref{sec:generation}).
(3) An \emph{Online Defense Generator}: it is a \textbf{dynamically linked library} that
(a) loads the patches from the configuration file $\mathcal{C}$ 
at program start, and (b) intercepts buffer allocation operations for
recognizing vulnerable buffers and generate security measures online (Section~\ref{sec:deployment}).

\subsection{Calling-Context Sensitive Patches} 

Given the attack input that exploits a heap vulnerability $\mathcal{V}$,
in order to generate a patch $\mathcal{P}$  based on attack 
analysis, it is critical to
extract some \emph{invariant} among attack
instances. Such invariant then can be used to design protection 
against future attacks that also exploit $\mathcal{V}$.

Our observation is that attacks that exploit $\mathcal{V}$
usually share some attack-time calling context (e.g., the sequence of
active function calls that lead to a buffer
overflow due to a \texttt{memcpy} call). If we trace the program execution backward,
these vulnerable buffers probably share 
the allocation-time calling context, which we call a \emph{vulnerable calling context} and 
can be used as an invariant to generate the patch $\mathcal{P}$. 
Whenever a heap buffer is allocated, the current calling
context is retrieved and compared with vulnerable calling contexts to determine whether
the buffer being allocated is vulnerable.

%% file: design_implementation.tex
\section{Targeted Calling Context Encoding} \label{sec:targeted}
Simple call stack walking for retrieving calling contexts would
incur a large overhead, especially for programs with
intensive heap allocations~\cite{pcc}.
There exist several efficient calling context encoding techniques that are famous, 
such as~\cite{pcc, pcce, deltapath}. 
We propose \emph{targeted calling context encoding}, which
is a suite of algorithms that can be used to
optimize these encoding techniques. The insight is that when the \emph{target functions}, whose
calling contexts are of interest, are known, many irrelevant call sites do not
need to be instrumented and thus the overhead can be
significantly reduced. 

The input of our algorithms is
the call graph of the program, and the output is
a pruned sub-call-graph. As each of the three
famous encoding techniques~\cite{pcc, pcce, deltapath}
can take a graph as input, they all
should work well when the sub-call-graph
if fed to these systems.  To make the discussion
concrete (and based on our choice of the encoding technique for heap patching),
we use \emph{Probabilistic Calling Context} (PCC)~\cite{pcc}
to demonstrate the application of the proposed optimizations.

According to PCC, at the prologue 
of each function, the current calling context ID (CCID),
which is stored in a thread-local integer variable $V$, is read
into a local variable $t$; right before each call site,
$V$ is updated as $V = 3*t + c$,
where $c$ is a random constant unique for each call site.\footnote{The encoding in PCCE~\cite{pcce}
and DeltaPath~\cite{deltapath} basically adopts $V = t + c$, where $c$ is calculated according
to the encoding algorithms.}  
This way, $V$ continuously stores the current CCID. 
Thus, the current CCID can be obtained conveniently by reading $V$.
With PCC, however,
it may happen that multiple calling contexts obtain the same
encoding due to hash collisions. 
It is shown  
practically and theoretically that the chance of hash collision is very low \cite{pcc}.
It is worth noting that a hash collision in our system means that
a non-vulnerable buffer is recognized as a vulnerable buffer and gets
enhanced. Any of our enhancements do not change the program
logic, so \textbf{a hash collision can cause unnecessary overhead,
but it does not affect the correctness of our system.}

\begin{figure}
    \centering
   \includegraphics[scale=.31]{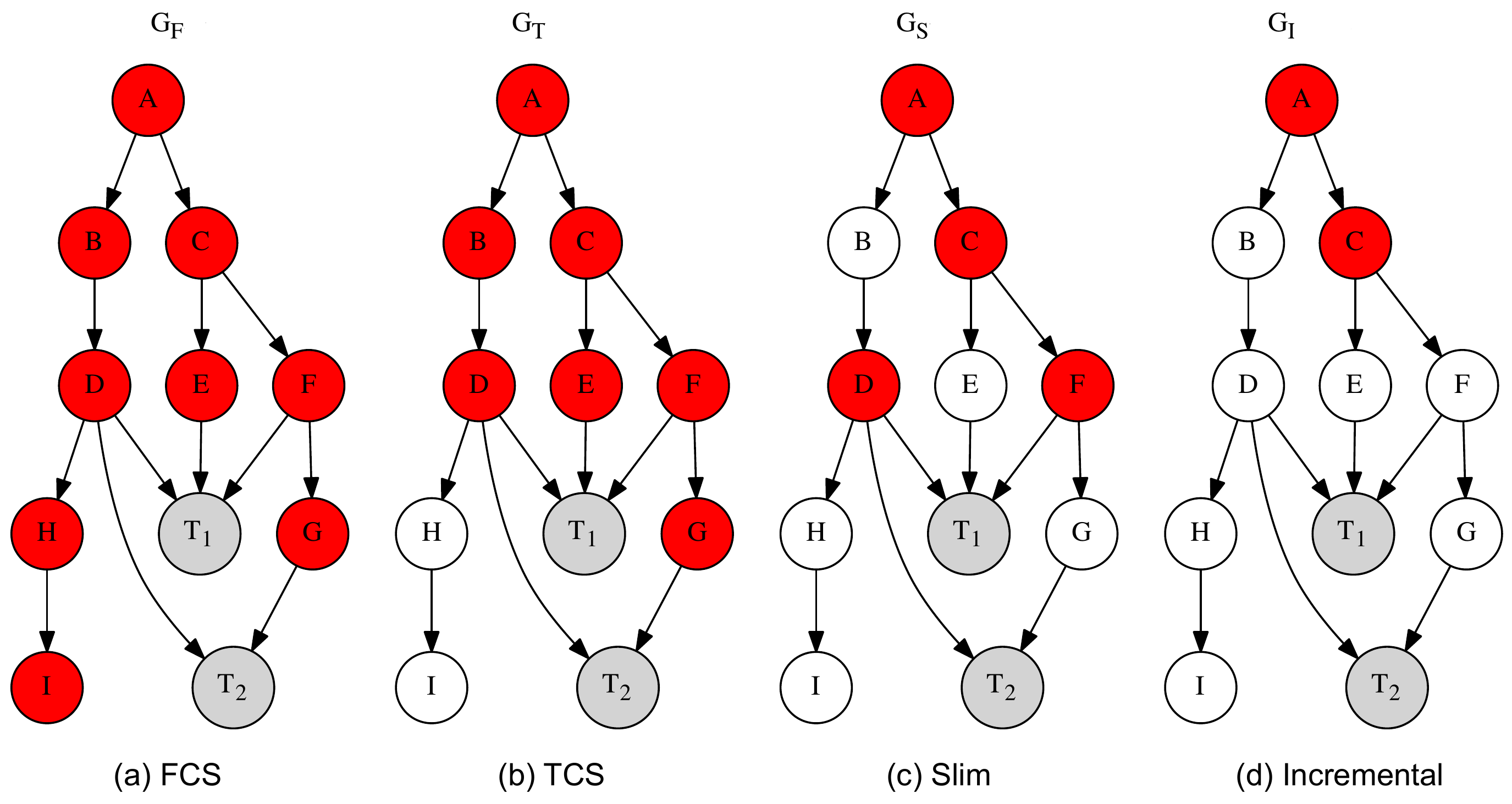}
    \caption{Comparison of different encoding optimization algorithms using an example call graph. The gray nodes, $T_1$ and $T_2$, are target functions; red nodes indicate functions whose call sites are instrumented; and white nodes indicate functions whose call sites are not instrumented. For the simplicity of presentation, the example does not include \emph{\textbf{back edges}}. Our algorithms can handle back edges without problems as shown in Algorithm~\ref{alg:incremental}.}
    \label{fig:pcc-summary}
\af
\end{figure}

We call the original encoding algorithm that take all the call sites into consideration
as Full-Call-Site (FCS) instrumentation. The three famous encoding
algorithms, PCC~\cite{pcc}, 
PCCE~\cite{pcce} and DeltaPath~\cite{deltapath}
all enforce FCS. Figure~\ref{fig:pcc-summary}(a)
shows that all the call sites in those red nodes are instrumented,
and $T_1$ and $T_2$ are the target functions. The less call
sites are instrumented, the smaller overhead is expected.

\subsection{Targeted-Call-Site (TCS) Optimization}
FCS blindly instruments all the call sites in a program.
In practice, very often users are only interested in the calling 
contexts that end at one of a specific set of target functions, such
as security-sensitive system calls
and critical transaction calls.
In our case, we are only interested in
calling contexts when the allocation APIs (such
as \texttt{malloc}, \texttt{calloc}, \texttt{calloc},
\texttt{memalign}, \texttt{aligned\_alloc}) are
invoked. It is unnecessary to instrument functions
that may never appear in the call stacks when these target functions
are invoked. 

We thus propose the first optimization, \emph{Targeted-Call-Site} (TCS), where
only the call sites that may appear in the calling
contexts of target functions are instrumented.  
To conduct the TCS optimization, reachability analysis on the call graph of the
program is performed. Given a call graph $G=\langle V, E \rangle$,
where $V$ is the set of nodes representing functions of the program
and $E$ the set of function calls, and a set of functions $\mathcal{F}$,
we perform reachability analysis to find edges that can reach any
of the functions in $\mathcal{F}$, and only call sites corresponding
to these edges are instrumented. 

Figure~\ref{fig:pcc-summary}(b) shows the instrumentation
result of TCS. As the edges $DH$ and $HI$ cannot reach
any of the target functions $T_1$ and $T_2$, they are 
pruned from the instrumentation, reducing
the set of call sites that need to be instrumented.

\subsection{Slim Optimization}
On the basis of TCS, there is still potential 
to further prune the set of call sites to be instrumented. 
In a call graph, a node can be classified as either a \emph{branching} or \emph{non-branching} one:
a branching node is one that has multiple outgoing edges that can reach
(one of) the target functions. \emph{Our insight} is that
the purpose of call site
instrumentation is to make sure different calling contexts can
obtain different encoding values;
given a non-branching node, whether or how its contained call sites are instrumented
does not affect the \emph{distinguishability} of the encoding results. 
Thus, we propose to avoid instrumenting the call sites in those
non-branching nodes.

For example, as shown in Figure~\ref{fig:pcc-summary}(c), 
according to the Slim optimization, all call sites
in the non-branching nodes, $B$ and $E$,
are excluded from the instrumentation set. 

\subsection{Incremental Optimization}
The two optimization algorithms treat all target functions as a whole.
Our another \emph{insight} is that when the call to a target function
is intercepted for analysis or logging purpose, the analyzer or
logger usually knows the target function.
In our case, when \texttt{malloc} and \texttt{memalign} are 
intercepted, different interception functions will be invoked.

Therefore, we can use the pair of $\langle$ \texttt{Target\_fun, CCID} $\rangle$
(rather than \texttt{CCID} alone) to distinguish different calling contexts.
Based on this insight, we propose another optimization algorithm that
can further reduce the number of instrumented call sites.
A node is an \emph{true branching node} if it has two or more 
outgoing edges that reach the same target function. That is,
if a node has multiple outgoing edges, each of which reaches a different target
function, it is called a \emph{false branching node}.
The idea of the Incremental encoding is to avoid instrumentation the
call sites in a false branching node.

In Figure~\ref{fig:pcc-summary}, node $A$ is a true
branching node, as its two outgoing edges can reach the
same node $T_1$ (and $T_2$ as well). So is node $C$,
as its two outgoing edges can reach
$T_1$. Thus, only the call sites that correspond 
to $AB$, $AC$, $CE$, $CF$ need to be instrumented. 
Take the calling contexts of $T_2$ as an example,
the instrumentation at $AB$ and $AC$ is sufficient
to distinguish the two calling contexts that reach $T_2$.

\begin{algorithm}
	\small
	\caption{\small Incremental Optimization.}
	\label{alg:incremental}

	\begin{algorithmic}[1]
		\INPUT A call graph $CG = \langle N, E \rangle$, and the set of target functions $ T \subseteq N $.
		\OUTPUT The functions in $ N $ to be instrumented.
		\Function{Filter}{$T , CG = \langle N, E\rangle$}:
			\State $ InstrumentationSet \leftarrow \lbrace \rbrace $

			\For{$ t \in T $}
				\State $ VisitedNodes \leftarrow \lbrace \rbrace $
			
				\State $Queue.push(t)$
			
				\For{$ n \leftarrow Queue.pop() $}
					\State $ VisitedNodes.push(n) $
					\For{\textbf{each} $ e = \langle m, n \rangle $ of the incoming edges of $n$}
						\If{$ m \notin VisitedNodes$ }
							\State $ Queue.push(m) $
						\EndIf
					\EndFor
				\EndFor
			
				\For{$ n \in VisitedNodes $} 
					\State $count \leftarrow 0$
					\For{\textbf{each} $ e = \langle n, m\rangle $ of outgoing edges of $n$}
						\If{$ m \in VisitedNodes $}
							\State $ count \leftarrow count + 1 $
						\EndIf
					\EndFor
					\If {$ count > 1 $}
						\State $ InstrumentationSet.push(n) $
					\EndIf
				\EndFor
			\EndFor
			
			\State \Return{$ InstrumentationSet $}
		\EndFunction

	\end{algorithmic}
\end{algorithm}

Algorithm~\ref{alg:incremental} shows the algorithm for incremental optimization.
Line 3 illustrates the idea of processing each target function incrementally. 
For each target function, Lines 4--17 are to find true branching nodes relative to it.
Specifically, Lines 4--10 are a backward breadth-first search; as it omits nodes
already visited (Line 9), it can correctly handle \emph{back edges}. 
Then Lines 11--17 are to find true branching nodes.

In short, the three encoding optimization algorithms 
are based on different insights and ideas, and each
improves on the previous one in terms of reducing the set of
call sites to be instrumented.

\begin{figure} 
    \centering
\aaf
   \includegraphics[width=0.51\textwidth]{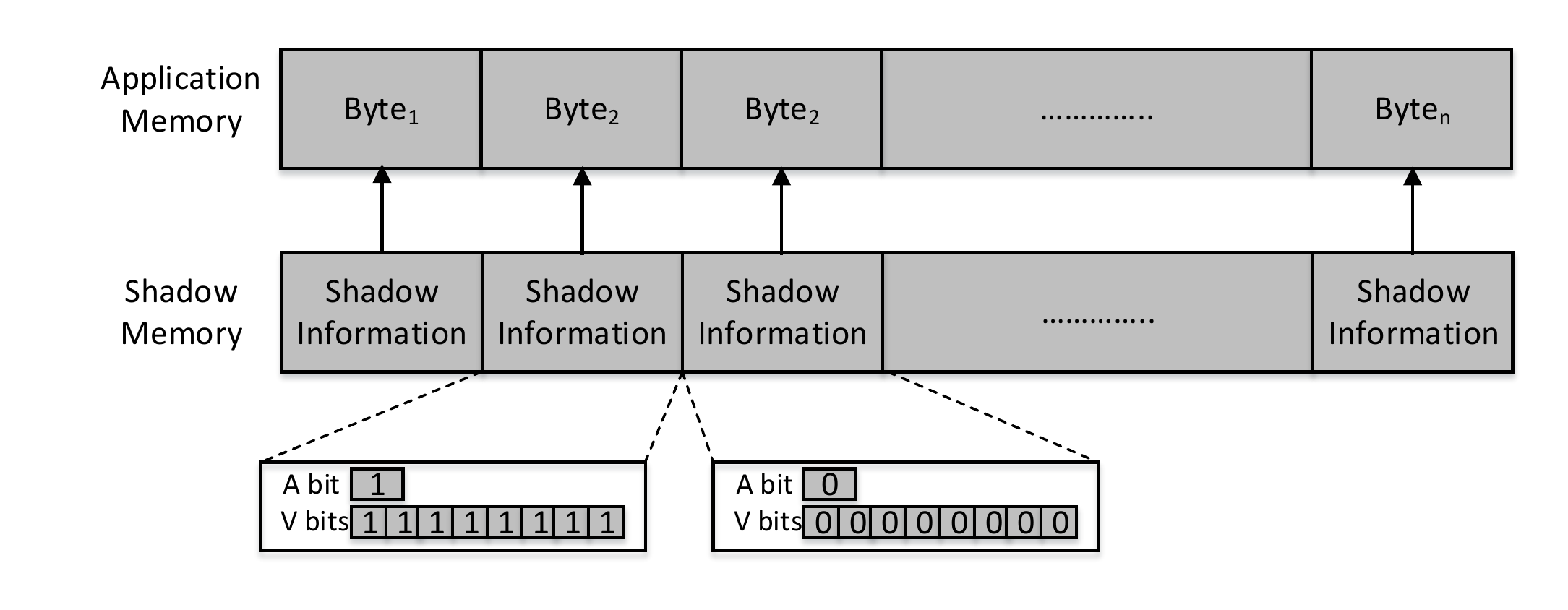}
\aaf    
    \caption{Shadow memory.}
    \label{fig:shadow_mem}
\af
\end{figure}

\section{Offline Attack Analysis and Patch Generation} \label{sec:generation}
The Offline Patch Generator component runs the vulnerable program
using the attack input and generates the patch as part
of the dynamic analysis report. It
is built on dynamic binary instrumentation and shadow 
memory of Valgrind~\cite{memcheck-shadow-memory}.
As shown in Figure~\ref{fig:shadow_mem},
for every bit of the program memory, a \emph{Validity} bit (V-bit) 
is maintained to indicate whether the accompanying bit has a 
legitimate value; instructions are inserted for the propagation
of V-bits when data copy occurs (e.g., when a word is read from
memory to a register);
for every byte of the memory location, an \emph{Accessibility} 
bit (A-bit) is maintained to indicate whether the memory 
location can be accessed. 

When a heap buffer is \emph{malloc}-ed, the returned memory is marked
as accessible but invalid. Each buffer is surrounded by a pair of 
\emph{red zones} (16 bytes each), which are marked as inaccessible. 
 When a heap buffer is \emph{free}-ed, its memory is 
set as inaccessible. In addition, whenever a heap buffer is allocated, 
the current calling context ID (CCID) is recorded and associated with the buffer. 

\newspace
\noindent \textbf{(1) Detecting overflows:} A buffer overflow will access the 
inaccessible red zone appended to the buffer and get detected. 

\newspace
\noindent \textbf{(2) Detecting use after free:} A \emph{free}-ed buffer is set as inaccessible and then added to a FIFO queue of freed blocks. Thus, the memory is not
immediately made available for reuse. Any attempts to access any of the
blocks in the queue can be detected. The maximum total size of the buffers
in the queue is set as 2GB by default, which is large enough for the exploits
we investigated, and can be customized. In Section~\ref{sec:discussion},
we discuss how to handle it if the quota is insufficient. 

\begin{figure}
    \centering
\begin{lstlisting}
typedef struct {
	uint32_t i; 
	uint8_t c;
} A;
A y, *p = (A *) malloc( sizeof(A) );
p->i = 0; p->c = 'f';
y = *p;
\end{lstlisting}
\aaf
    \caption{Legal uninitialized read due to padding.}
    \label{fig:padding}
\af
\end{figure}

\newspace
\noindent \textbf{(3) Detecting uninitialized read:} 
To detect uninitialized read, an attempt is to
report any access to uninitialized data, but this will lead 
to many false positives. For instance, given the code snippet in Figure~\ref{fig:padding}, most
of the compilers will round the size of \texttt{A} to 8 bytes;
so only 5 bytes of the heap buffer is initialized (and the V-bits for
the remaining 3 bytes are zero), while the
compiler typically generates code to copy all 8 bytes for 
\texttt{y = *p}, which would cause false positives due to accessing
the 3 bytes whose V-bits are zero.

\begin{figure}
\centering
\aaf
   \includegraphics[scale=.5]{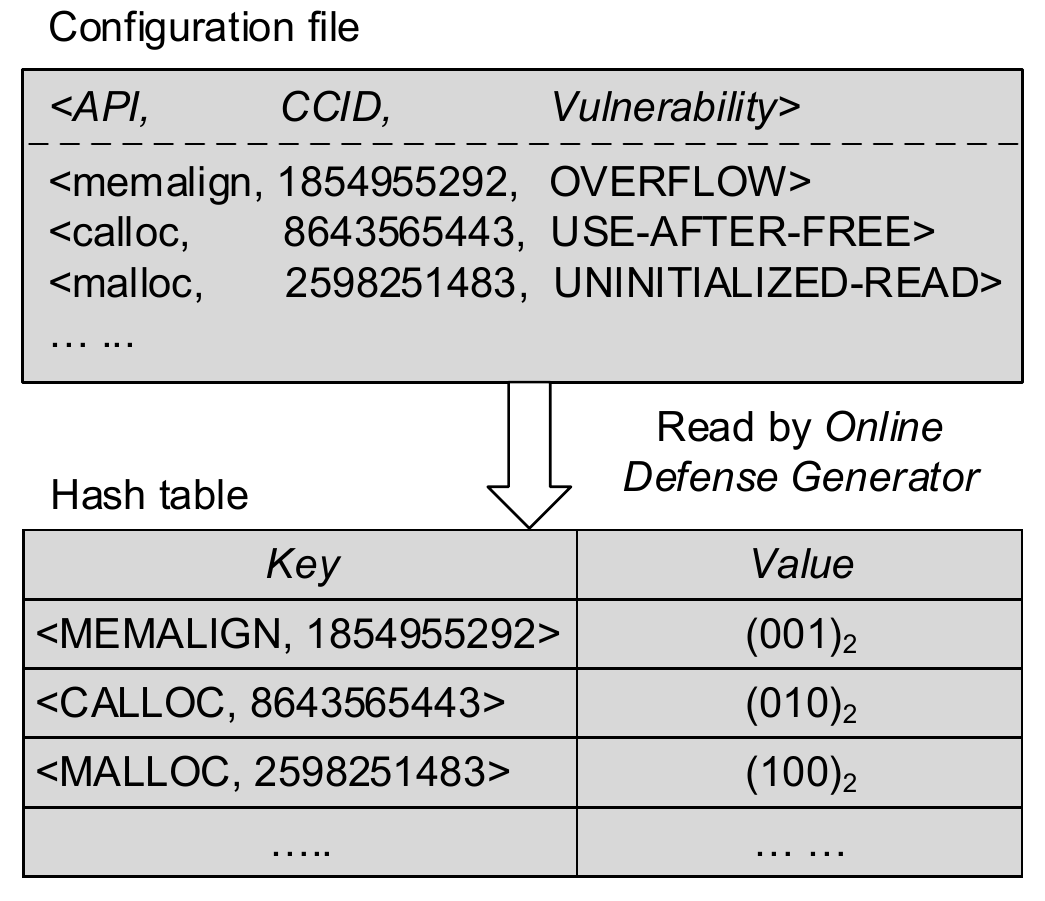}
    \caption{Patches read into a hash table.}
    \label{fig:hashtable}
\af
\end{figure}

To avoid false positives due to padding,
we check the V-bit of a value only when it is used 
to decide the control flow (e.g., \texttt{jnz}), 
used as a memory address, or used in a system
call (as the kernel behavior is not tracked). 
As every bit of the program has a V-bit, 
bit-precision detection of uninitialized read is achieved. 
Moreover, \emph{origin tracking} is used to
track the use of invalid data back to the uninitialized data (such
as a heap buffer) when a warning is raised, which allows us
to retrieve the allocation-time CCID associated with
the vulnerable buffer. 
When an attack is detected, the patch is generated in the form
of $\langle$ \texttt{FUN, CCID, T}$\rangle$, where \texttt{FUN} is the function
used to request the heap buffer (such as \texttt{malloc}, \texttt{memalign}),
\texttt{CCID} is an integer representing
the allocation-time calling context ID of the vulnerable buffer,
and \texttt{T} is a three-bit integer representing
the vulnerability type 
(the three bits are used to indicate OVERFLOW, USE-AFTER-FREE, 
UNINITIALIZED-READ, respectively).
Example patches are shown in the upper graph in Figure~\ref{fig:hashtable}.

\noindent \textbf{How to handle \texttt{realloc}:}
If the new size is smaller than the original size,
the cut-off region is marked as inaccessible.
If the new size is larger, the added region 
is set as accessible but invalid. The allocation-time
CCID associated with the buffer is also updated
with the value upon the \texttt{realloc} invocation.

\noindent \textbf{How to handle multiple vulnerabilities:} 
An attack input may exploit multiple vulnerabilities.
For example, the Heartbleed attack exploits both uninitialized
read and overread. In order to handle the case
that an attack exploits multiple vulnerabilities, 
we resume the program execution upon warnings.
Plus, once the V bits for a value have been 
checked, they are then set to \emph{valid}; this avoids 
a large number of chained warnings.
Finally, a script is used to process the many
warnings according to the origin (i.e., the address of
the vulnerable buffer) of those warnings and generate
patches correctly.

\section{Code-less Patching and Online Defenses} \label{sec:deployment}
When the patched program is started, as shown in Figure~\ref{fig:hashtable},
the \emph{Online Defense Generator} library has an initialization function\footnote{\texttt{\_\_attribute\_\_((constructor))} is used to declare the function.}
that reads patches 
from the configuration file and
stores them into a \emph{hash table}, where the key of each entry is 
$\langle$ \texttt{ALLOCATION\_FUNCTION}, \texttt{CCID}$\rangle$ and the value
is the vulnerability type(s) and parameters, if any, for applying
the security measures. 
\textbf{Note once the hash table is initialized, its memory pages
are set as read only.}

The Online Defense Generator library intercepts all
heap memory allocation operations. Whenever a heap
buffer is allocated, the name of the allocation
function (hardcoded into the interposing function) 
along with the current CCID is used
to search in the patch hash table, which takes
only O(1) time. If there is no match, the buffer
does not need to be enhanced; otherwise,
the buffer is enhanced based on the associated
vulnerability type(s) and parameters. 

While the security measures themselves are
straightforward, several considerations make
the design  challenging. (1) In some
cases, the same buffer may be vulnerable to
multiple attacks, such as uninitialized read
and overflow.
(2) In addition to handling \texttt{malloc}
and \texttt{free}, the system needs to support
a family of other allocation functions, such
as \texttt{realloc} and \texttt{memalign} (aligned
allocation). These challenges are 
well resolved by our system.

One complexity is that we maintain
heap metadata ourselves, such as the buffer size 
(to support \texttt{realloc} correctly),
vulnerability type(s), the buffer alignment
information, and the location of the guard page,
so that \textbf{our system can work
without having to change the underlying allocator
or rely on its internals.}

\begin{figure} 
    \hspace{-7pt}
   \includegraphics[scale=0.39]{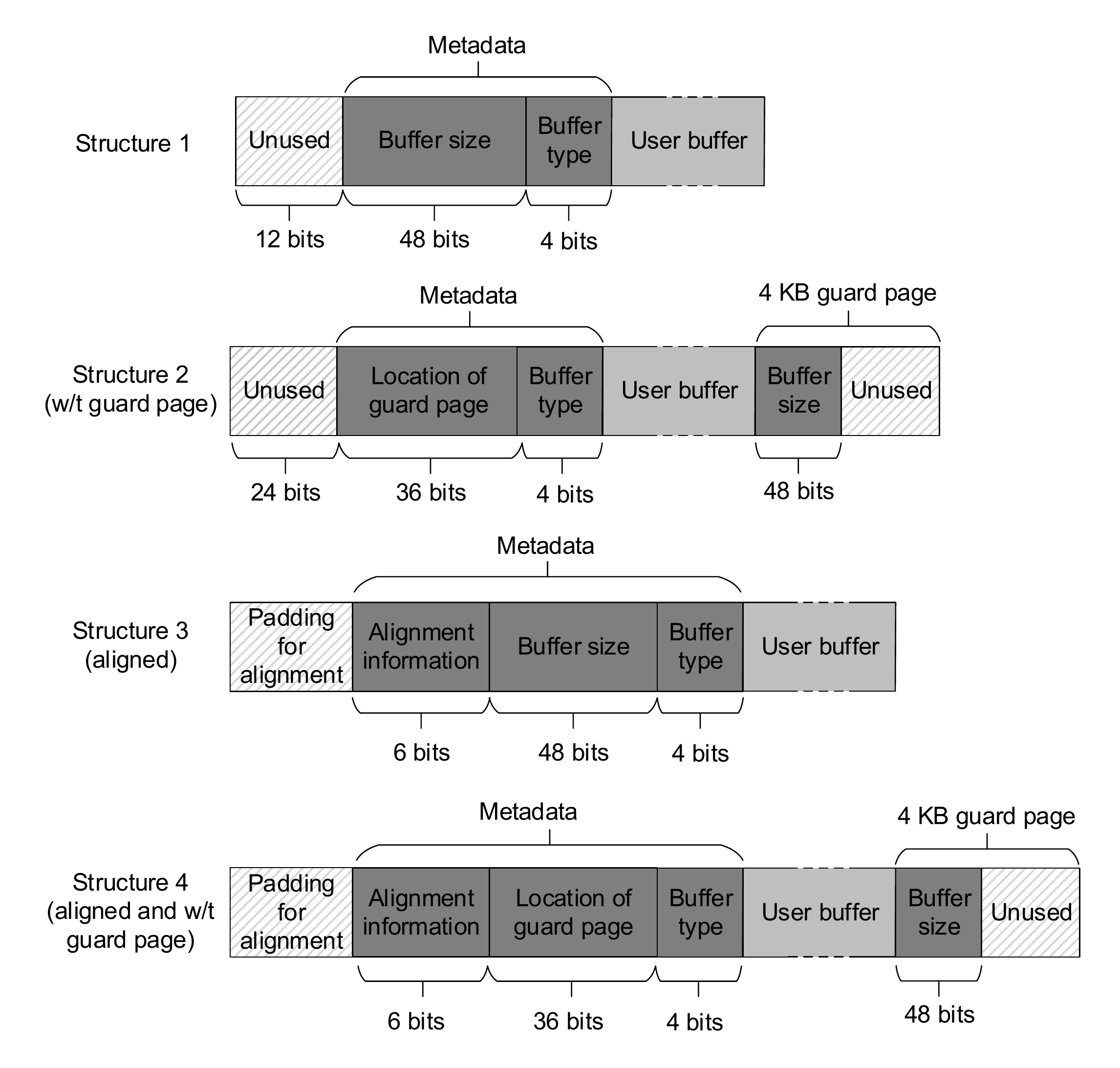}
  \aaf
    \caption{Buffer structures. Note how we pack the metadata into
    only one word (64 bits) preceding the \emph{user buffer}.}
    \label{fig:buf_struc}
\af
\end{figure}

\newspace
\noindent \textbf{(1) Handling overflows:} If the 
buffer is vulnerable to overflows, a guard
page is appended to it to prevent such attacks. 
While the guard page can effectively prevent
overflows, they are known to be
prohibitively expensive when being applied to
every buffer. In our system, however,
the guard page is precisely applied to vulnerable
buffers, and the resulting overhead is dramatically
reduced.

As shown in Figure~\ref{fig:buf_struc}, Structure 2 is used
for non-aligned buffers, while Structure 4 is used
for aligned buffers (allocated using \texttt{memalign}, etc.).
When a heap allocation request is intercepted, the
requested size is increased to accommodate the word
for \emph{metadata} and the guard page (as well as necessary 
padding following the user buffer to ensure the guard page
is page aligned).
The address of the user buffer is returned to service the user program.

The \emph{metadata} word contains rich information and
is worth detailed interpretation. 
(1) In all structures, the least significant four bits
is called the \emph{buffer type} field, where three bits represent
the \emph{vulnerability type} (one bit is used to indicate each
of the three vulnerability types, i.e., \texttt{Overflow}, \texttt{Use
after Free}, and \texttt{Uninitialized Read}) and one bit indicates
whether the buffer is \emph{aligned}. 
(2) 36 bits are used to indicate the location of
the guard page. Currently, 64-bit operations systems 
only use a 48-bit virtual address space; plus, a guard page
is 4KB=$2^{12}B$ aligned. Thus, $48-12=36$ bits are sufficient. 
A guard page is set as inaccessible
using \texttt{mprotect}. 
The user buffer size information is stored as the first word
of the guard page, and it is needed for supporting
\texttt{realloc}.
(3) If the buffer is aligned (Structure 3 and Structure 4),
there is a padding field whose size depends on the alignment size.
The alignment size information is needed to 
determine the buffer address given the address of the User Buffer upon a \texttt{free} call.
As the alignment size is always a power of two (i.e., $2^n$),
we only need  6 bits to store the value of $n \in [0, 64]$, which
then can be used to calculate the alignment size.

\newspace
\noindent \textbf{(2) Handling use after free:} 
If an allocation is not aligned, the buffer takes
Structure 1; otherwise, Structure 3. 
The metadata word uses
48 bits to store the user buffer size. 
When a buffer vulnerable to use after free is 
to be \emph{free}-ed, it is put into an FIFO queue of freed blocks
to defer the reuse.
In our system,
only buffers vulnerable to use-after-free
are put into the queue, such that given the same quota 
the time a freed buffer stays in the queue
is much lengthened, which hence significantly increases
the difficulty of exploitation
of a use-after-free vulnerability for it increases
the uncertainty entropy a freed buffer is reused
by attackers. 

\newspace
\noindent \textbf{(3) Handling uninitialized read:} 
Similar to the above, if 
the allocation is not aligned, the buffer takes
Structure 1; otherwise, Structure 3. 
The \emph{user buffer} region is initialized
with zeros before it is returned to the user program.

\begin {table}
\caption {A summary of the use of buffer structures.}
\label{tab:buf_struc_use}
\centering
 \begin{tabular}{|c | c | c| } 
 \hline
 \textbf{Vulnerability type} & \textbf{Not aligned} & \textbf{Aligned} \\ 
 \hline
 Not Vulnerable & Structure 1 & Structure 3\\
 \hline
 Overflow & Structure 2 & Structure 4 \\ 
 \hline
 Use-after-free & Structure 1 &  Structure 3 \\
 \hline
 Uninitialized read & Structure 1 & Structure 3 \\
 \hline
\makecell{Overflow \& \\  Use-after-free } & Structure 2  & Structure 4\\
 \hline
 \makecell{Overflow \& \\ Uninitialized read } & Structure 2 & Structure 4 \\ 
 \hline
 \makecell{ Use-after-free  \& \\ Uninitialized read } & Structure 1 & Structure 3\\
 \hline
 \makecell{Overflow \& \\  Use-after-free  \& \\ Uninitialized read } & Structure 2 &  Structure 4 \\ [1ex] 
 \hline
\end{tabular}

\end {table}

Table~\ref{tab:buf_struc_use} summarizes how
different buffer structures are used for handling different
cases, including when multiple vulnerabilities affect the same buffer.
If there is a threat of overflow, Structure 2
 or Structure 4 is used to accommodate the guard page
depending on whether the allocation call is \texttt{memalign}.
Whenever there is use after free, upon being freed the buffer is put into
the freed-blocks queue to defer the reuse of these buffers.

\begin{figure}
    \centering
   \includegraphics[scale=0.53]{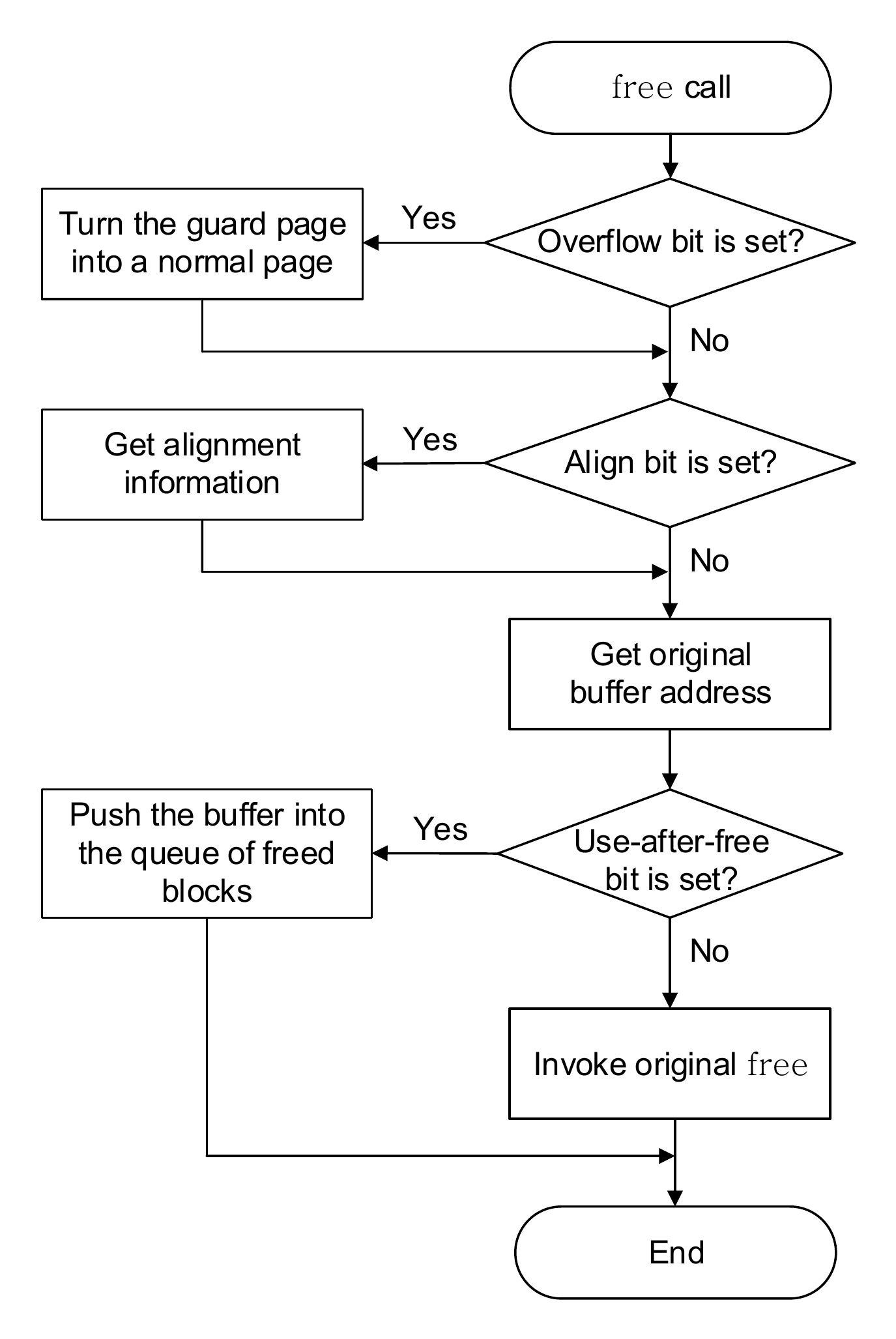}
    \caption{Handling \texttt{free()}.}
    \label{fig:free_flow}
\end{figure}

\newspace
\noindent \textbf{How the Online System Handles \texttt{free()} calls:}
A particular advantage of our system is that it supports 
the deployment of heap patches without modifying
the underlying allocator. 
It works solely by intercepting the memory allocation calls. 
On the other, it complicates the handling of freeing buffers.

As shown in Figure~\ref{fig:free_flow}, when \texttt{free(p)} is invoked by the user program,
the Online Defense Generator intercepts the call and handles
it as follows. (1) If the Overflow bit in the metadata word
is set, the location information of the guard page is retrieved and 
the guard page is set as accessible using \texttt{mprotect}. 
(2) Based on the user buffer address \texttt{p}, 
the initial address of the buffer \texttt{pi} is calculated. Specifically,
if the buffer was not allocated using \texttt{memalign},
\texttt{pi = p - sizeof(void*)};
otherwise, the alignment size \texttt{A} is retrieved and
\texttt{pi = p - A}. 
(3) If the Use-after-Free bit is set, the block is put
into the queue of the freed blocked; otherwise, the buffer
is released using the original \texttt{free} API of the
underlying allocator and the buffer address \texttt{pi}
is passed to the call.

%% file: implementation.tex
\section{Other Implementation Details}
\newspace 
\noindent \textbf{Implementation of the Program Instrumentation Tool.}
Currently, we add a pass into LLVM, which performs
the call graph analysis to determine the set of call sites to be
instrumented and then instruments them. This implementation assumes
the program source code is available. Another viable implementation path is based
on binary code instrumentation, e.g., via Dyninst, which can insert code 
into the program dynamically.\footnote{So far, our 
Dyninst-based solution only supports single threaded programs. } 
It is worth mentioning that Dyninst does not require the source
code to perform instrumentation. Thus, software users (not only software companies) 
can also instrument their software (only once) and generate patches themselves.

\newspace 
\noindent \textbf{Implementation of the Offline Patch Generator.}
This component is built on the basis of Valgrind~\cite{memcheck-shadow-memory}. We reuse its 
shadow memory functionality and modify the tool to support the
needed handling of allocation and deallocation. Significant effort
has been saved by making use of Valgrind, which in the meanwhile
is a mature dynamic analysis tool. The implementation over Valgrind
also benefits us to analyze various complex real-world programs successfully.

\newspace 
\noindent \textbf{Implementation of the Online Defense Generator.}
It is implemented as a shared library, which reads
the patches in the configuration file to the hash table and
interposes all the allocation function calls to enhance vulnerable buffers
according to the patches. The hash table memory pages are set as read-only once
the initialization is done.
The library is loaded using \texttt{LD\_PRELOAD} in our prototype. 
It does not change the underlying heap allocator
or rely on its internals.

%% file: evaluation.tex
\section{Evaluation} \label{eval}
We have evaluated \textsc{HPaC} in terms of
both effectiveness and efficiency. 
We not only evaluate it on the SPEC CPU2006 benchmarks and
many vulnerable programs,
but also run the system with real-world service 
programs. The efficiency improvement of
the calling context encoding optimization
algorithms is also measured. 
Our experiments 
use a machine
with a 2.8GHZ CPU, 16G RAM running 16.04 Ubuntu and Linux Kernel 4.10.

\begin{center}
\begin {table} 
\caption {Vulnerable programs used in the evaluation. \texttt{UR} and 
\texttt{RaF} stand for uninitialized read and use after free, respectively.
}
\label{tab:effect_table}
 \begin{tabular}{c c c} 
 \hline
 \textbf{Program} & \textbf{Vulnerability} & \textbf{Reference}\\
 \hline
 Heartbleed & UR \& Overflow & CVE-2014-0160\\

 bc-1.06 & Overflow & Bugbench\cite{Lu05bugbench:benchmarks}\\ 

 GhostXPS 9.21 & UR & CVE-2017-9740\\

 optipng-0.6.4 & UaF & CVE-2015-7801\\

 tiff-4.0.8 & Overflow & CVE-2017-9935\\

 wavpack-5.1.0 & UaF & CVE-2018-7253\\ 

 libming-0.4.8 & Overflow & CVE-2018-7877\\

SAMATE Dataset & Variety & 23 heap bugs\cite{samate}\\
 \hline
\end{tabular}
\af
\end {table}
\end{center}

\subsection{Effectiveness} \label{effect}
To evaluate the effectiveness of our system \textsc{HPaC},
we run it on a series of programs, as shown in Table~\ref{tab:effect_table}, 
which contain a variety of heap vulnerabilities. In the effectiveness
experiments, we aim to evaluate 
(1) whether the Offline Patch Generator can correctly
determine the vulnerability type and generate patches; and
(2) whether the generated patches 
can effectively prevent attacks from exploiting those heap vulnerabilities.
It is worth mentioning
that \textsc{HPaC}, as a single system, can handle the variety of heap
vulnerabilities; plus, different from conventional 
code patches, installing patches do not
modify any line of the program code and hence do not
introduce stability or logic errors (but we
do require one-time program instrumentation). 
Below we describe details for some of the experiments
we performed. 

\newspace
\noindent \textbf{Heartbleed Attacks.} Heartbleed was a notorious vulnerability
of \texttt{OpenSSL} and affected a large number of services~\cite{Durumeric:2014:MH:2663716.2663755}. By sending an ill formed heartbeat request
and receiving the response, the attacker can steal data
from the vulnerable services, such as private keys and user
account information. While Heartbleed is widely known as 
a heap buffer over-read vulnerability, actually the
attacker can exploit two different vulnerabilities:
over-read and uninitialized read. Specifically, the
vulnerable heap buffer has 34KB, while the size $l$ of the
data stealing from the buffer can be up to 64KB. 
If $l < 34K$, the attack is just an uninitialized read that
leaks data previously stored  in the buffer;
otherwise, it is a mix of uninitialized read and over-read~\cite{Wang2015a}.

A service was created using the \textit{OpenSSL} utility 
\texttt{s\_server}.\footnote{In order to support the interposition of the allocation
operations, we compiled OpenSSL using 
\texttt{OPENSSL\_NO\_BUF\_FREELIST} 
compilation flag to disable the use of freelists.} 
We then collected different attack inputs from Internet, and used one of
them to generate the patch. Our Offline Patch Generator correctly
identified it as a mix of uninitialized read and overflow and output
the patch. The patch was then automatically written into the configuration
file of the Online Defense Generator, which was able to  precisely recognize and enhance
the vulnerable buffers. We then tried different attack 
inputs, and no data was leaked except for the zeros filled in the buffers.

\newspace
\noindent \textbf{bc-1.06.} \textit{bc}, for basic calculator, is an arbitrary-precision 
calculator language with syntax similar to the C programming language. 
Some versions of its implementation contain a heap buffer overflow vulnerability.  
We obtained a buggy version of this program from BugBench, a C/C++ bug benchmark 
suite~\cite{Lu05bugbench:benchmarks}, and collected a malicious input that overflows
buffers and corrupts the adjacent data. By feeding the input into our Offline Patch Generator 
that ran the buggy program, an overflow patch was generated. 
With the patch deployed, our system successfully stopped the attack 
before it corrupted any data.

\newspace
\noindent \textbf{GhostXPS 9.21.} \texttt{GhostXPS} is an implementation of the Microsoft 
XPS document format built on top of \texttt{Ghostscript}, which is an interpreter/renderer 
for PostScript and normalizing PDF files. It is the leading independent interpreter software 
with the most comprehensive set of page description languages on the market today.  
Some versions of \texttt{GhostXPS} contain an uninitialized read vulnerability 
that can be exploited using a crafted document. We collected a buggy version of \textit{GhostXPS} 
from their git repository and the malicious document input. 
In the offline patch generation phase, the uninitialized read attack was detected and 
a patch was generated. During the online heap protection phase,
the attack was not able to steal any data, except for zeros, 
from memory.

\newspace
\noindent \textbf{optipng-0.6.4.} \texttt{OptiPNG} is a \texttt{PNG} image 
optimizer that compresses image files to a smaller size without losing any 
information. Specific versions of this optimizer allow the attacker to 
exploit a use-after-free vulnerability and execute arbitrary code via  
crafted PNG files. We collected a vulnerable version (\texttt{optipng-0.6.4}) 
and a malicious \textit{PNG} image. The Offline Patch Generator correctly 
identified the attack and generated a patch. The Online Defense Generator made 
use of the patch to recognize the vulnerable buffers and defeated the 
use-after-free attacks by deferring the deallocation of vulnerable buffers.

\newspace
\noindent \textbf{tiff-4.0.8.} \texttt{TIFF} provides support for "Tag Image File Format", 
commonly used for sorting image data. In \texttt{LibTIFF 4.0.8}, there is a heap buffer 
overflow in the \texttt{t2p\_write\_pdf} function in \texttt{tools/tiff-2pdf.c}.
We were able to generate the patch, which could successfully prevent the 
overflow.

\newspace
\noindent \textbf{SAMATE Dataset.} 
We evaluated our system on the \texttt{SAMATE Dataset}, which is maintained by
NIST~\cite{samate} and contains 23
programs with heap buffer overflow, uninitialized read, or use after free vulnerabilities.
Our system successfully generated patches for all of them and 
prevented the vulnerabilities from being exploited.

\subsection{Efficiency} \label{efficiency}
We compared the overhead incurred by the different calling context encoding
algorithms, and measured the overall speed overhead and memory overhead incurred by our system.
We used our LLVM-based implementation to measure the efficiency of different
calling context encoding algorithms.

\subsubsection{Overhead Comparison of Different Calling Context Encoding Algorithms}
To measure the execution time overhead imposed by different calling
context encoding algorithms, we applied them to the programs in
the SPEC CPU2006 Integer benchmarks, and measured the execution time when
different encoding techniques were applied, normalized using
the execution time when no encoding is applied. 
Compared to FCS (Full Call-Site Instrumentation) proposed
in ~\cite{pcc}, which incurred 2.4\% of slowdown
for C/C++ programs, the
other three encoding algorithms proposed by us, that is, 
TCS (Targeted Call-Site Instrumentation), Slim, and Incremental, 
incurred 0.6\%, 0.5\%, and 0.4\% of slowdown, receptively. 
While the saved execution time itself is small, it gains up to 6x of speed up. 
We believe the proposed encoding algorithms can have many
applications far beyond memory protection; plus, when they
are applied to Java programs, where FCS may incur more than
35\% of overhead~\cite{deltapath}, the speed up due to our algorithms
could make a significant difference. 

\begin{center}
\begin {table}
\caption {SPEC CPU2006 benchmark program size increase, in percentage, 
due to different encoding algorithms.}
\label{tab:encoding_binary_size}
 \begin{tabular}{l    r  r  r  r } 
 \hline
 Benachmark &  FCS(\%) & TCS(\%) & Slim(\%) & Incremental(\%) \\ [0.5ex] 
 \hline
 400.perlbench & 19.6 & 16.2 & 	15.9 &	15.9 \\

 401.bzip2 & 8.8 &	0.12 &	0.12 &	0.12 \\

 403.gcc &	18.6 & 14.7 &	13.6 &	13.6	 \\ 

 429.mcf &	0.53 &	0.53 &	0.53 &	0.53\\

 445.gobmk & 4.8 &	3.2 &	2.5 &	2.5\\

 456.hmmer & 18.9 &	5.9 & 2.4 &	1.2\\

 458.sjeng & 10.6 & 0.08 &	0.08 &	0.08\\ 

 462.libquantum	& 15 &	7.7 &	7.7 &	7.7\\

 464.h264ref & 8.3 &	3.6 & 	1.8 &	1.8\\

 471.omnetpp & 15.8 &	7.2 & 6.7 &	6.7\\ 

 473.astar & 7.0 & 7.0 & 0.2 & 	0.2\\

 483.xalancbmk & 14.5 &	4.1 &	3.8	& 3.8\\ [1ex] 
 \hline
\end{tabular}
\af
\end {table}
\end{center}

As the encoding works by inserting instructions into the
programs, we also measured the program size increase.
The results are shown in Table~\ref{tab:encoding_binary_size}.
While FCS increased the binary size by an average of 12\% when
compared to the uninstrumented binaries, TCS, Slim and
Incremental incurred only 6\%, 4.5\%, and 4.4\% of size increase, 
respectively.

\subsubsection{Efficiency of \textsc{HPaC}}
To evaluate the run-time overhead of our system, 
we ran our system on both SPEC CPU2006 Integer benchmarks
and a set of real-world service programs. 

\newspace
\noindent \textbf{SPEC CPU2006.} 
The speed overhead incurred by \textsc{HPaC} can be divided
into four parts: (1) overhead due to instrumentation,
which has been presented above; (2) overhead due to 
interposition of heap memory allocation calls; 
(3) overhead due to maintaining the meta data
of each buffer (such that our system does not
rely on the internal details of the underlying allocator);
(4) overhead due to patch deployment, which causes
the security measures to be applied to vulnerable buffers . 

In order to measure the overhead incurred due to
patch deployment, we select a set of allocation-time CCIDs (Calling-Context IDs) as 
hypothesized vulnerable ones as follows. First,
for each benchmark program, we rank all of its allocation-time
CCIDs according to their frequencies during the profiling execution (that is,
how many heap buffers have been allocated
under that calling context). Next, we pick the CCIDs
with \emph{median} frequencies as the hypothesized vulnerable ones.
Finally, we regard the heap buffers with those allocation-time CCIDs
as ones vulnerable to overflows (the other two vulnerability
types are much less expensive to treat), and generate
corresponding patches for them. 

\begin{figure*}[t]
    \centering
   \includegraphics[width=0.8\textwidth]{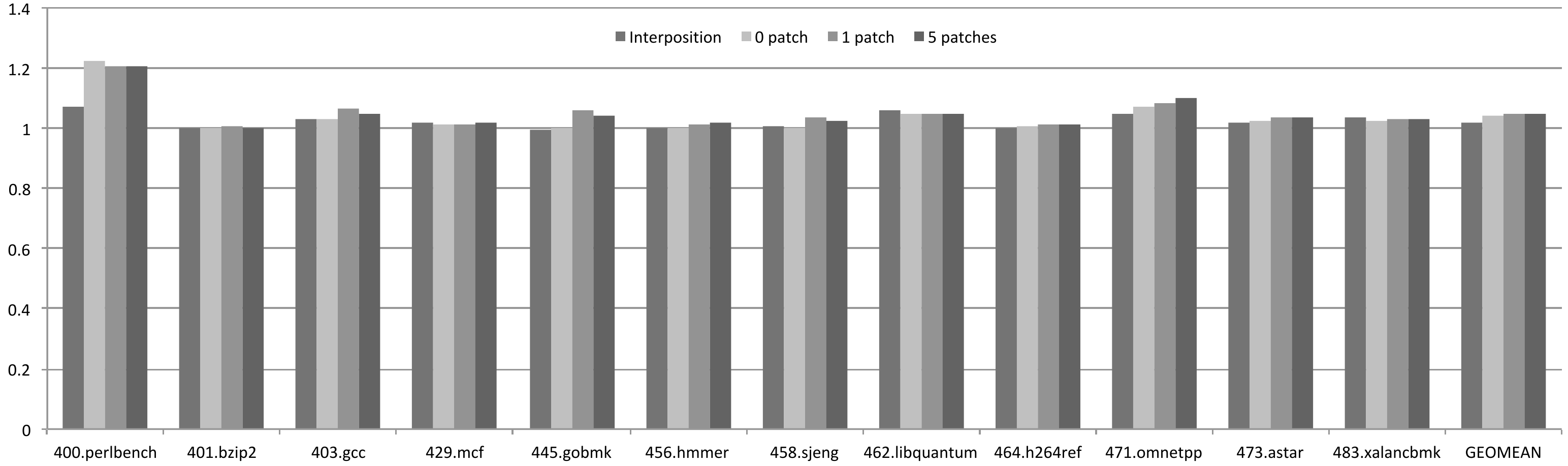}
    
    \caption{Execution time overhead imposed to SPEC CPU2006 benchmarks, when only interposition is applied, maintaining
    the meta data for each buffer (plus the interposition overhead), one patch is installed, and five patches are installed,
    respectively.}
    \label{fig:runtime} 
\end{figure*}

Figure \ref{fig:runtime} shows the measurement results.
The  overhead due to interposition is 1.9\%, and
the overhead for maintaining the meta data  for
buffers (plus the interposition overhead) is 4.3\%.
Note that this part of overhead can be easily eliminated
if our system is integrated with the underlying heap allocator.
When one patch is installed, the overhead becomes 4.7\%, only 0.4\% of
overhead increase. 
The total overhead is 5.2\% when five patches are installed.
One outlier is \texttt{400.perlbench}, which has the most
intensive heap allocations. Table~\ref{tab:malloc-count} records the heap allocation statistics
for each SPEC CPU2006 benchmark. 

\begin{center}
\begin {table}
\centering
\caption {SPEC CPU2006 benchmark heap allocation statistics.}
\label{tab:malloc-count}
 \begin{tabular}{l    r  r  r  r } 
 \hline
 Benachmark &  malloc & calloc & realloc\\ [0.5ex] 
 \hline
 400.perlbench & 346,405,116 & 0 &  11,736,402\\

 401.bzip2 & 174 &	0 &	0\\

 403.gcc &	23,690,559 & 4,723,237 &	44,688	 \\ 

 429.mcf &	5 &	3 &	0 \\

 445.gobmk &  606,463 &	0 &	52,115\\

 456.hmmer &  1,983,014 &	122,564 &  368,696 \\

 458.sjeng & 5 & 0 & 0\\ 

 462.libquantum	& 1 &	121 &	58\\

 464.h264ref & 7,270 &	170,518 & 	0 \\

 471.omnetpp &  267,064,936 &	0 &	0\\ 

 473.astar &  4,799,959  & 0 & 0 \\

 483.xalancbmk &  135,155,553 &	0 &	0\\ [1ex] 
 \hline
\end{tabular}
\af
\end {table}
\end{center}

\begin{figure}
    \centering
   \includegraphics[width=0.48\textwidth]{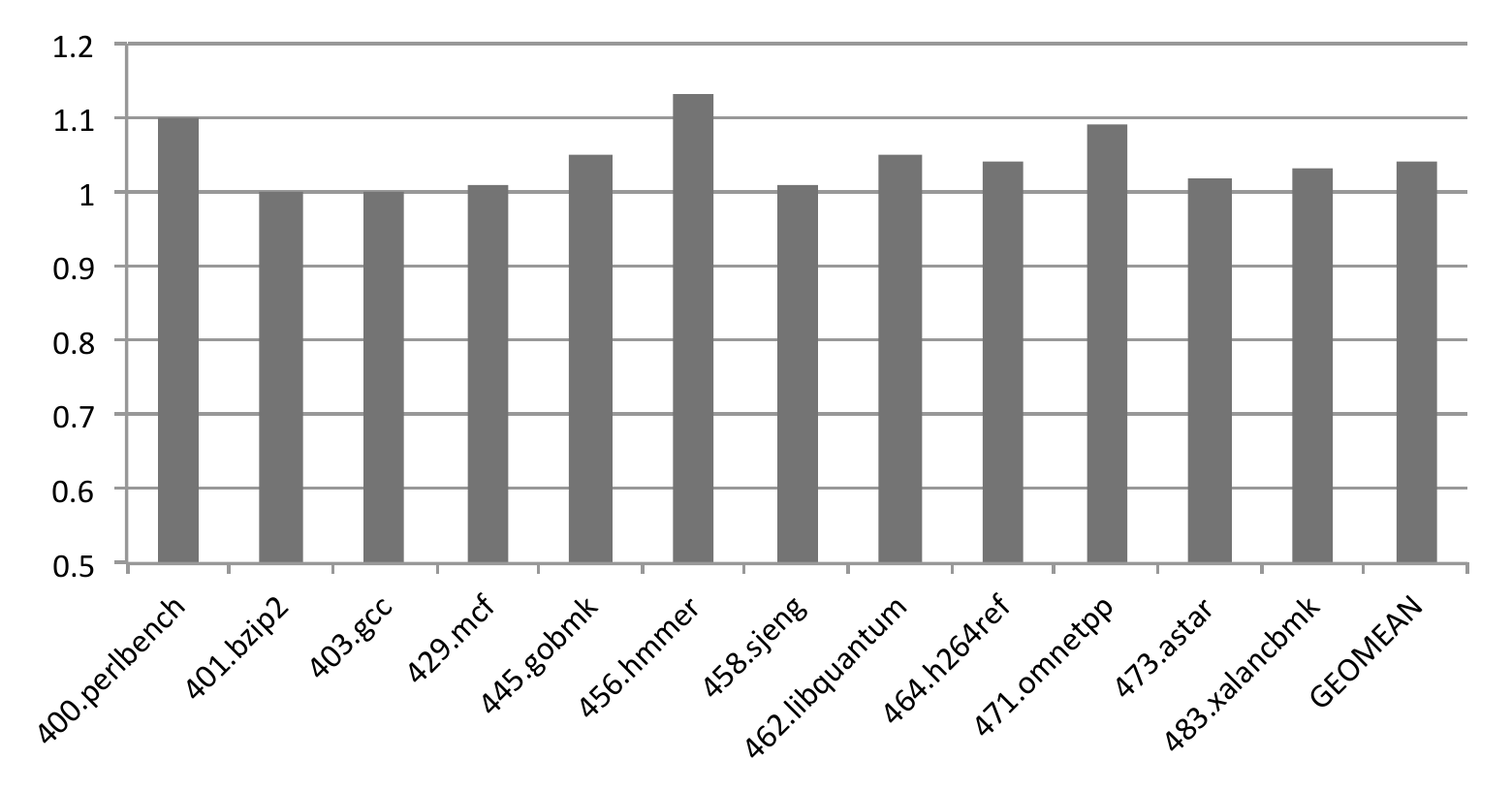}
    
    \caption{Normalized memory overhead imposed to SPEC CPU2006 benchmarks when \textsc{HPaC} runs.}
    \label{fig:memory_overhead}
\end{figure}

We also measured the memory consumption of benchmark programs with and without
our system deployed. To perform the experiments we used a script that can compute the memory overhead in terms of the average Resident Set Size (RSS) for the benchmark programs. That script reads the \texttt{VmRSS} field of \texttt{/proc/[pid]/status}. The sampling rate of that script is 30 times per second; then, the mean of the reading is calculated. Figure~\ref{fig:memory_overhead} shows
the memory consumption overhead normalized over native program execution, and the average memory overhead is only 4.3\%.

\newspace
\noindent \textbf{Service Programs.}
We also evaluated our system on two popular service programs:
\texttt{Nginx} and \texttt{MySQL}. 
We used \texttt{Nginx} 1.2, and measured the throughput 
overhead by sending requests
using \texttt{Apache Benchmark}. Different numbers of concurrent
requests were used, and the throughput was compared with that
of native execution. The average throughput overhead is
only 4.25\%. 

\texttt{MySQL} 5.5.9 was used, and we applied the built-in
test script to measuring the throughput overhead. There was no
observable throughput overhead. The memory overhead
in both cases was negligible. Note that the memory overhead
is proportional with the number of \emph{live} heap buffers.

\newspace
\noindent \textbf{Summary.}
The evaluation shows that \textsc{HPaC} is not only effective
but also very efficient. Its most optimized calling context
encoding incurs only 0.4\% of slowdown,
a 6 times of speed boost compared to the original  encoding algorithm. 
The 4.3\% speed slowdown, due to allocation/deallocation interposition 
and metadata maintaining,
 can be largely eliminated if our system is integrated with the underlying heap allocator. 
Installing a single hypothesized heap patch only incurs another 0.4\% of speed
slowdown. It is noticeable the throughput overhead
on real-world service programs is very low or negligible. 

Unlike many heap protection systems that typically incur
a very high overhead (e.g., 2.5X of slowdown using MemorySanitizer~\cite{stepanov2015memorysanitizer} that detects
deterministic uninitialized read detection only and 20\% of
slowdown using Dieharder~\cite{Novark:2010:DSH:1866307.1866371} that provides probabilistic protection) 
and/or handles only one heap vulnerability
type (e.g., HeapTherapy~\cite{Zeng2015}), \textsc{HPaC} handles multiple frequently exploited
heap vulnerability types with high efficiency. 
The patch generation is precise and automatic, and 
the patch deployment does not require manual intervention and
does not modify the program code, guaranteeing that no new bugs
are generated.

%% file: limitations.tex
\section{Discussion amd Future Work} \label{sec:discussion}
A limitation is that \textsc{HPaC} can only handle buffer overflows due 
to continual write or read operations, which are the main form of buffer overflows. 
Overflows due to discrete read or write  cannot be handled by \textsc{HPaC}. Plus, 
if an overflow runs over an array which is an internal field of a buffer, \textsc{HPaC} cannot detect it. 
The limitation is common in many existing countermeasures against buffer overflows, such
as AddressSanitizer~\cite{addresssanitizer} and Exterminator~\cite{novark2008exterminator}.

It may happen that a heap vulnerability can be exploited with multiple CCIDs, 
and thus the attacker may develop different attack input to exploit buffers 
with new allocation calling contexts. However, whenever the attack exploits a buffer allocated in a new calling context, 
our system simply treats it as a new vulnerability and starts another defense cycle. 
More importantly, based on our evaluation and previous researches on 
context-sensitive defenses~\cite{automatic, rx, Zeng2015, novark2008exterminator}, this is rare.

We do not claim that \textsc{HPaC} is to replace
existing patching system. It is to complement
conventional patching by providing immediate and bug-free protection when the fresh
patches are still not mature and may need more time for testing.

For programs that have a large memory profile, to analyze the use-after-free attack,
the memory quota for the FIFO queue of freed blocks may be exceeded. 
In this case, we can replay attacks in multiple
executions; specifically, we divide the whole space of CCIDs
into $N$ subspaces, and each of the $N$ executions defers the deallocation of buffers
that have the allocation-time CCIDs in one of the subspaces.
Now, each execution is expected to consume $1/N$ of the
memory. 

Some works aim to discover more zero-day heap vulnerabilities before 
software release~\cite{dowser,borg,203682}, while
other works try to pinpoint memory defects by analyzing the 
core dumps~\cite{Xu:2016:CTL:2976749.2978340,
203880}. How to combine them with our system
for proactive patch generation and
patch generation without attack replay is an interesting problem;
we will explore it as our future work. 

%% file: conclusions.tex
\section{Conclusions}
We have combined heavyweight offline attack analysis,
lightweight online defense generation, and program instrumentation
to build a new heap memory defense system \textsc{HPaC}.
It has overcome the challenge of generating online defenses
for handling a variety of vulnerabilities, and even \emph{combo}
vulnerabilities (a buffer that can be exploited by different types
of attacks). The task is further complicated when buffer metadata
maintainence is transparent to the underlying heap allocators, and does
not assume any special custom allocators. 

The new heap defense system has many prominent advantages:
(1) \emph{patch generation without
manual efforts}, 
(2) \emph{code-less patching},
(3) \emph{versatile} handling of heap buffer overwrite, overread, use after free, and
uninitialized read,
(4) \emph{imposing a very small overhead}, and
(5) \emph{no dependency on specific allocators}.
In addition, we have proposed \emph{targeted calling context encoding},
which should interest researchers applying or building 
calling context encoding techniques. 
The evaluation shows that the system as well as the encoding optimization is effective
and efficient. The speed overhead is only 5.2\% when five patches are
installed on SPEC CPU2006 benchmarks.